\documentclass{PoS}
\def\Dslash{\mathop{\not\!\! D}}

\title{Pseudoscalar flavor-singlets and staggered fermions}

\ShortTitle{Pseudoscalar flavour-singlets and staggered fermions}

\author{\speaker{Eric B. Gregory}, Alan C. Irving, Christopher M. Richards\\
        Theoretical Physics Division, Department of Mathematical Sciences,
	University of Liverpool, Liverpool L69-7ZL, United Kingdom\\
        E-mail: \email{egregory@liverpool.ac.uk},\hspace{2mm}
	\email{aci@liv.ac.uk},\hspace{2mm}
	\email{cmr@liv.ac.uk}}

\author{Craig McNeile\\
 Department of Physics and Astronomy, The Kelvin Building,
       University of Glasgow, Glasgow G12-8QQ, United Kingdom\\
        E-mail: \email{c.mcneile@physics.gla.ac.uk}}

\abstract{The Asqtad improved staggered fermion formalism has
been a valuable tool in successfully calculating the non-singlet parts of
the hadronic spectrum. We are engaged in a project to calculate
the spectrum of the pseudoscalar singlet mesons with $2+1$-flavor Asqtad 
staggered gauge configurations. Propagators of flavor-singlet 
states incorporate contributions from both disconnected and connected diagrams,
and hence are sensitive to any differences in the actions governing the sea 
and valence fermions on the lattice. As such, they also present the 
possibility of 
a probe of the validity of the ``fourth-root trick'' in the staggered fermion 
formulation.  We present an update on our progress toward measuring the 
$\eta'$ mass on $2+1$-flavor Asqtad staggered gauge configurations, including
a review of methods and preliminary results. We also show a strong 
correlation between ${\rm Tr}(\gamma_5\otimes 1)$ and the topological charge 
in these configurations, as predicted by the index theorem.}

\FullConference{XXIV International Symposium on Lattice Field Theory\\
		 July 23-28 2006\\
		 Tucson Arizona, US}

\begin{document}

\section{Introduction}
  The pseudoscalar flavor-singlet meson system is of theoretical interest for
a variety  of reasons. It differs from the non-singlet by the inclusion of
disconnected diagrams in its propagator, to which is attributed the mass 
difference between the pion and the $\eta'$ meson 
(958 MeV)\cite{Witten:1979vv,Veneziano:1979ec}.
 A thorough lattice QCD calculation of the spectrum of the pseudoscalar 
flavor-singlet system is needed for full comprehension of this connection. 
To date there have been a number of lattice studies of $N_f=2$ flavor singlet 
mesons\cite{Itoh:1987iy,McNeile:2000hf,Struckmann:2000bt,Lesk:2002gd,
Schilling:2004kg,DeGrand:2002gm,Venkataraman:1997xi,Kogut:1998rh,
Fukaya:2004kp}.
These proceedings contain a report on an $N_f=2+1$ flavor Wilson lattices 
fermion study from CPPACS \cite{:2006xk}.

We are engaged in a study of the pseudoscalar singlet system with $N_f=2+1$ 
flavors of improved staggered quarks. The motivation is two-fold. First,
the improved staggered ``Asqtad'' formulation has a formidable track record of 
accurate reproduction and prediction of experimentally measurable quantities,
in part due to the relative high speed of staggered fermion simulation and the
accessibility of relatively light quark masses\cite{Aubin:2004wf}. 
Second, the staggered 
formulation has been plagued by theoretical questions regarding the validity
of the ``fourth-root trick'' employed in simulations where $N_f\neq 4$. 
These concerns center on whether or not the ${\rm det}^{1/4}$ of the native
four-flavor staggered fermionic matrix introduces pathologies into the sea
quarks. An additional motivation, then, is to see whether singlet propagators,
which are uniquely sensitive to the fermionic sea, illuminate any such 
pathologies.

For $N_f$ degenerate flavors of quarks, the full propagator is
\begin{equation}
G_{\eta'}(x',x)=\langle\sum_{i}\overline{q}_i(x')(\gamma_5\otimes{\bf 1})q_i(x')\sum_{j}\overline{q}_j(x)(\gamma_5\otimes{\bf 1})q_j(x)\rangle,
\end{equation}
where the $\gamma_5\otimes{\bf 1}$ denotes that the meson has $\gamma_5$ Dirac 
spinor structure, and is a singlet in staggered ``taste'' space. This 
expression includes $N_f$ connected terms (pion propagators):
\begin{equation}
\langle\sum_{i}\overbrace{\overline{q}_i(x')(\gamma_5\otimes{\bf 1})\underbrace{q_i(x')\sum_{j}\overline{q}_j}(x)(\gamma_5\otimes{\bf 1})q_j}(x)\rangle,
\end{equation}
and $N_f^2$ disconnected terms:
\begin{equation}
\langle\sum_{i}\overbrace{\overline{q}_i(x')(\gamma_5\otimes{\bf 1})q_i}(x')\sum_{j}\overbrace{\overline{q}_j(x)(\gamma_5\otimes{\bf 1})q_j}(x)\rangle.
\end{equation}
In other words
\begin{equation}
G_{\eta'}(x',x)= N_fC(x',x) - N_f^2D(x',x),
\end{equation}
where the additional fermion loop in each of the disconnected diagrams gives rise to the negative sign.

As the connected correlators are the propagators for the non-singlet meson we 
expect, when the ground state dominates, for them to decay exponentially in time separation  
on a Euclidean lattice:
\begin{equation}
G_{\pi}(t)=C(t)\sim e^{-m_\pi t}.
\end{equation}
 Furthermore,
\begin{equation}
 G_{\eta'}(t)=N_fC(t)-N_f^2D(t)\sim e^{-m_{\eta'}t}.
\end{equation}
Then the ratio of disconnected to connected contributions should be
\begin{equation}
\label{general_ratio}
R(t)=\frac{N_f^2D(t)}{N_fC(t)}= 1-Ae^{-(m_{\eta'}-m_\pi)t},
\end{equation}
if the sea quarks are appropriately dynamical. In the case of quenched QCD we expect 
instead \cite{Bernard:1992mk,Bernard:1993sv}
\begin{equation}
\label{quenched_R}
R(t)\sim A + Bt
\end{equation}
It is also possible that the ${\rm det}^{1/4}$ in the staggered formulation 
introduces other pathologies that may manifest themselves in this ratio. 

Measuring disconnected correlators is a difficult task. They are noisier than
connected correlators, as they contain fluctuations of the fermionic sea. 
Hence one must use some sort of all-to-all propagators to get as much 
information out of a particular gauge configuration as possible. Additionally, 
the pseudoscalar singlet is closely related to very slow topological modes, 
necessitating the use of very long time series. We will describe how we address
these challenges below.

\section{Simulation and Measurement}
We have at our disposal the large publicly-available library of MILC Asqtad
lattices. Our preliminary study has predominately utilized the ``coarse
lattice'' ensembles, with lattices of $20^3\times 64$ and lattice spacing of 
$a \approx 0.125{\rm fm}$\cite{Bernard:2001av}. The typical ensemble has 
between 400 to 600 configurations separated by 6 trajectories.

On each gauge configuration we measure the connected correlators with standard 
point sources with routines integrated into the Chroma code 
\cite{Edwards:2004sx}. We measure disconnected correlators with stochastic 
volume-filling sources\cite{Markham:2000br,Farchioni:2004ej,Foley:2005ac}, 
$\eta(x)$. Given that on averaging over sources
\begin{equation}
\delta_{x,x'}=\langle\eta(x)^\dagger\eta(x)\rangle_\eta,
\end{equation}
and defining $M\phi=\eta$, then the pseudoscalar loop operator is
\begin{equation}
\label{operator}
{\mathcal O}_{\gamma_5\otimes {\bf 1}}(t)=\sum_{x_4=t}
\langle\phi^\dagger(x)\Delta_{\gamma_5\otimes{\bf 1}}\eta(x)\rangle_\eta.
\end{equation}
The operator $\Delta_{\gamma_5\otimes{\bf 1}}$ is the appropriate four-link covariant shift, displacing the quark and antiquark to opposite corners of the 
hypercube, with appropriate Kogut-Susskind phasing to effect the 
$\gamma_5\otimes{\bf 1}$ operator. We average over $N_s=64$ noise sources. The 
disconnected correlator is then:
\begin{equation}
D(t)=\frac{1}{T}\sum_{t'}{\mathcal O}(t'){\mathcal O}(t'+t)
\end{equation}

In practice one can do better than Equation \ref{operator}. Venkataraman and 
Kilcup\cite{Venkataraman:1997xi} describe a variance reduction (VKVR) trick 
applicable 
to the four-link pseudoscalar singlet operator. Recognizing that the Asqtad
$\Dslash$ connects only odd sites to even sites, and the $M^\dagger 
M= (D^2+ m^2)$ and its inverse connects even to even and odd to odd only then:
\begin{equation}
m\langle\phi^\dagger{\Delta_{\gamma_5\otimes{\bf 1}}}\phi\rangle = 
\langle\phi^\dagger{\Delta_{\gamma_5\otimes{\bf 1}}}\eta\rangle,
\end{equation}
with the left-hand expression having a minimized variance. The equivalence of 
the expectation value depends on the $\Delta_{\gamma_5\otimes{\bf 1}}$ 
operator separating the source and sink by an even number of links (four). 

The numerical work by the TrinLat group suggests that using a set of dilute 
noise sources, 
each defined on a subset of the lattice but collectively spanning the lattice, 
can reduce the variance of measured propagators\cite{Foley:2005ac}. In the case
of disconnected correlators for pseudoscalar singlets with Asqtad fermions, we 
found no discernible advantage.

\vspace{0.24in}
\begin{figure}[ptbh!]
\begin{center}
\scalebox{0.3}{\rotatebox{0}{\includegraphics{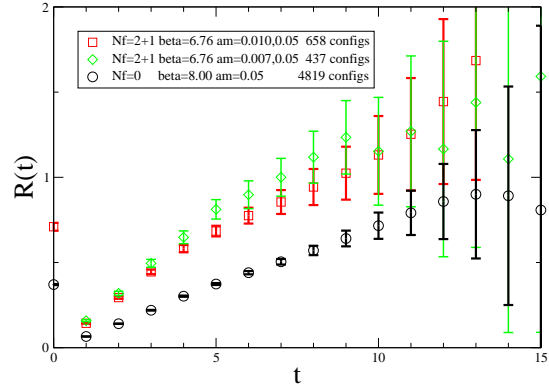}}}
\end{center}
\caption{\label{all_ratios} A comparison of $D/C$ ratios for quenched and 
$N_f=2+1$ flavor configurations. The $\beta=2.76$ $am=0.007$ points were 
computed with only 40 noise sources per configuration and without the VKVR
trick.
}
\end{figure}

\begin{figure}[ptbh!]
\begin{center}
\scalebox{0.3}{\rotatebox{0}{\includegraphics{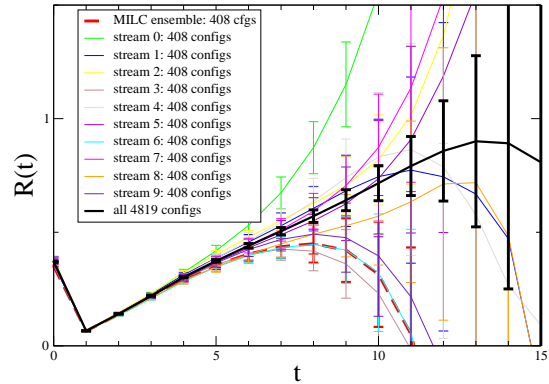}}}
\end{center}
\caption{\label{quenched_rat_binned} $D/C$ ratio for 4819 $\beta=8.00$, 
$am=0.05$, quenched configurations, and for subsets of 408 configurations,
including the original MILC ensemble.
}

\end{figure}
\begin{figure}[ptbh!]
\begin{center}
\scalebox{0.3}{\rotatebox{0}{\includegraphics{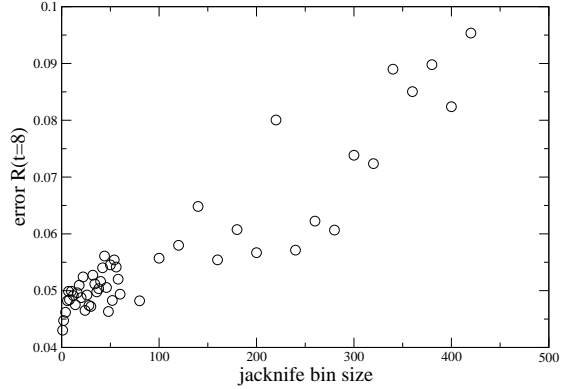}}}
\end{center}
\caption{\label{quenched_raterr_binsize} Error on $R(t=8)$ for 4819 
$\beta=8.00$, $am=0.05$, quenched configurations, as a function of jackknife 
bin size.
}
\end{figure}

\section{Analysis}
We were able to extract signals for connected and disconnected 
correlators for several ensembles of MILC coarse lattices. For $N_f=2+1$ 
configurations with two degenerate flavors of light sea quarks and one strange
sea quark flavor, expression \ref{general_ratio} generalizes 
to\cite{Gregory:2005me}
\begin{equation}
R(t)=\frac{4D_{qq}(t)+4D_{qs}(t)+D_{ss}(t)}{2C_{qq}(t)+C_{ss}(t)}.
\end{equation}
The disconnected correlators, $D_{qq}$, $D_{qs}$ and $D_{ss}$, are formed 
with two light quark loop operators, with a light loop operator and a 
strange loop operator, and with two strange loop operators, respectively. 
The connected correlators $C_{qq}$ and $C_{ss}$ are light quark and strange 
quark correlators, respectively.

When we form the
$D/C$ ratio for $N_f=2+1$ lattices we get a curve marginally consistent with 
a plateau at one as in Equation \ref{general_ratio}. However, it is also 
consistent with a linear relationship such as that in  Equation 
\ref{quenched_R}. With the MILC quenched ensemble ($\beta=8.0$ am=0.050, 408 
configs), we find a $D/C$ curve that is inconsistent with a linear 
relationship as in Equation \ref{quenched_R}. See Figure \ref{all_ratios}.

We hypothesized that the 
$\gamma_5\otimes{\bf 1}$ loop operator might be subject to the particularly 
long autocorrelation times of slow topological charge modes. As quenched 
configurations are relatively inexpensive to produce, we have extended the MILC
ensemble by nearly a factor of 12 to 4819 configurations. When we analyze the
entire ensemble, the $R(t)$ curve is more consistent with Equation 
\ref{quenched_R} (see figure  \ref{quenched_rat_binned}). However, subsets 
of 408 configurations are often 
mutually inconsistent by a margin of two standard deviations even when 
determined with relatively large jackknife bin size --- far greater than 
the integrated autocorrelation time of about 10 configurations. We interpret 
this to mean that the time series autocorrelation does not decay as a single 
exponential, and that there are autocorrelation time scales in this mode
that are as long as 100 configurations or more.

\section{Topological Charge}
\begin{figure}[ptbh!]
\begin{center}
\scalebox{0.3}{\rotatebox{270}{\includegraphics{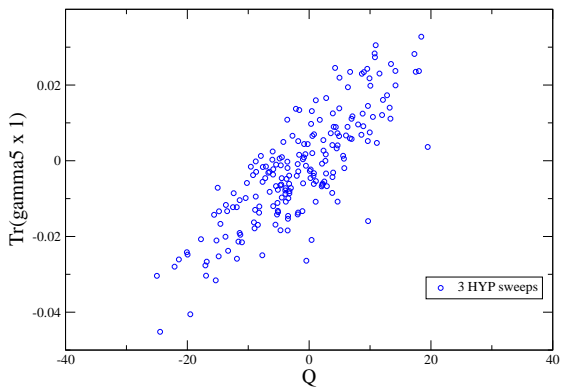}}}
\end{center}
\caption{\label{topolog_Q} Fermionic versus gluonic topological 
charge (after 
three HYP cooling sweeps) for $N_f=2+1$ $\beta=6.76$ $am=0.01,0.05$. 
Normalizations not consistent between the axes.
}
\end{figure}

  Using the Atiyah-Singer index theorem \cite{Atiyah:1968mp}, 
Smit and Vink \cite{Smit:1986fn} equate a fermionic expression for the 
topological charge 
\begin{equation}
\label{ferm_q}
Q=m\kappa_p\langle {\rm Tr}\left(\gamma_5 M^{-1}\right)\rangle_U
\end{equation}
with the traditional gluonic definition:
\begin{equation}
\label{glue_Q}
Q(x)=\frac{g^2}{64\pi^2}\epsilon^{\mu\nu\rho\sigma}F^a_{\mu\nu}(x)F^a_{\rho\sigma}(x)
\end{equation}
Equation \ref{ferm_q} is merely the loop operator in Equation \ref{operator}, 
integrated over time. 
 In \cite{Alles:1998jq}, Alles {\it et al.} show the equivalence of these two
definitions on Wilson configurations. As a cross check of our loop operator, 
we measured the topological charge using the gluonic definition, Equation 
\ref{glue_Q}, using hypercubic blocking\cite{Aubin:2004qz} implemented in the 
MILC code \cite{MILC_code}. We found a strong linear correlation between the 
integrated loop operator and the topological charge as defined gluonically in 
Equation \ref{glue_Q}. (See Figure \ref{topolog_Q} for an example.)

\section{Conclusions}
We interpret these results to mean that there is potential in Asqtad 
fermions as an avenue for investigating pseudoscalar singlet systems. 
If our experience with quenched configurations is a guide, one can achieve
good signals for disconnected propagators with the order of $10^4$ 
trajectories. While this is far in excess of any current MILC Asqtad
ensembles, we are currently in the process of generating such a
$\sim2\times 10^4$ trajectory ensemble on two racks of a QCDOC machine. 
Use of the VKVR trick has proven to be invaluable for reducing the variance
of disconnected $\gamma_5\otimes{\bf 1}$ operators. Careful measurement of 
a suite of fuzzed operators may also help in extracting the
pseudoscalar singlet system spectrum for $N_f=2+1$ staggered QCD, as well as 
shed light on the validity of the fourth-root trick.

\end{document}